\documentclass[preprint]{aastex}

\shorttitle{LBV in NGC 3432}
\shortauthors{Wagner et al.}

\begin{document}

\title{Discovery and Evolution of an Unusual Luminous Variable
Star in NGC
3432 (Supernova 2000ch)}

\author{R. M. Wagner}
\affil{Large Binocular Telescope Observatory, 933 North Cherry Avenue,
Tucson, AZ
85721}
\email{rmw@as.arizona.edu}

\author{F. J. Vrba, A. A. Henden, B. Canzian, \& C. B. Luginbuhl}
\affil{United States Naval Observatory, Flagstaff Station, Flagstaff,
AZ 86001}
\email{(fjv,aah,bc,cbl)@nofs.navy.mil}

\author{A.  V.  Filippenko, R.  Chornock, W. Li, \& A. L. Coil}
\affil{Department of Astronomy, University of California, Berkeley, CA
94720-3411}
\email{(alex,chornock,weidong,acoil)@astro.berkeley.edu}

\author{G.  D.  Schmidt \& P.  S.  Smith}
\affil{Steward Observatory, University of Arizona, Tucson, AZ 85721}
\email{(gschmidt,psmith)@as.arizona.edu}

\author{S. Starrfield}
\affil{Department of Physics and Astronomy, Arizona State University,
Tempe, AZ
85287}
\email{sumner.starrfield@asu.edu}

\author{S. Klose}
\affil{Th\"{u}ringer Landessternwarte Tautenburg, 07778 Tautenburg,
Germany}
\email{klose@tls-tautenburg.de}

\author{J. Tich\'{a} \& M. Tich\'{y}}
\affil{Klet Observatory, Zatkovo nabrezi 4, CZ-370 01 Ceske
Budejovice, Czech Republic}
\email{(jticha,mtichy)@klet.cz}

\newpage

\author{J. Gorosabel}
\affil{Laboratorio de Astrofísica Espacial y Física Fundamental, Apdo.
Correos 50.727, E-28080 Madrid, Spain}
\email{jgu@laeff.esa.es}

\and

\author{R. Hudec \& V. Simon}
\affil{Astronomical Institute of the Czech Academy of Sciences, 251 65
Ond\v{r}ejov, Czech Republic}
\email{(rhudec,simon)@asu.cas.cz}




\begin{abstract}

We present photometric and spectroscopic observations of SN 2000ch, an
unusual
and extremely luminous variable star located in the galaxy NGC 3432.
The
object was discovered on 2000 May 3.2 during the course of the Lick
Observatory
Supernova Search, at an unfiltered magnitude of about 17.4.
Pre-discovery
images obtained in 1997, 1998, and 2000 April show the object with $R
=
19.2-19.5$ mag.  Optical spectra obtained beginning on 2000 May 6 show
a
smooth, flat continuum and strong, broad hydrogen Balmer emission
lines at
wavelengths consistent with the catalogued redshift of NGC 3432,
strengthening
the association of the variable with the galaxy. Photometric
monitoring reveals
a complex and erratic light curve over a time span of $\sim$10 days.
Subsequent optical spectra over the next $\sim3$ months continued to
show
strong Balmer emission lines with a mean full-width at half-maximum
intensity
$\sim 1550$ km s$^{-1}$ and a distinct red asymmetry.  A spectrum
obtained 9
months after the outburst is similar to the previous spectra, but the
integrated flux in H$\alpha$ is nearly half that observed during the
outburst. The object's photometric behavior, spectrum, and luminosity
suggest
that it is a very massive and luminous variable star and might be
related to some luminous blue variable stars such as
$\eta$
Carinae and SN 1997bs in NGC 3627.  The brightest apparent magnitude
implies an
absolute magnitude of $M_V \approx -12.7$ at the distance of NGC 3432,
a value
which is comparable to $\eta$ Carinae during its outburst in the
mid-nineteenth
century.

\end{abstract}


\section{Introduction}

Recent systematic searches for supernovae (SNe) in external galaxies
such as
the Lick Observatory Supernova Search (LOSS; Li et al. 2000;
Filippenko et
al. 2001) have revealed not only a plethora of new SNe but many other
luminous
variable objects as well.  Some of these include a class of peculiar
type II
supernovae (SNe~II), designated SNe IIn, which exhibit hydrogen
emission lines
consisting of a broad complex base underlying a stronger, sometimes
slightly
blueshifted, narrow emission component with no trace of the broad
P-Cygni
profiles that are characteristic of normal SNe~II (Schlegel 1990;
Filippenko
1991).\footnote{Sometimes, however, a very narrow, low-velocity
absorption
component is visible; see SNe 1994W and 1994ak in Figure 14 of
Filippenko
(1997), for example.} SNe~IIn probably originate from the explosion of
massive
stars in which the high-speed ejecta and hard radiation field interact
with
the dense circumstellar envelope, created previously by extreme
mass-loss from
the progenitor.

It has become increasingly clear that some objects cataloged
previously as
peculiar SNe~II are actually SNe~IIn while others are not genuine SNe,
defined
to be the final demise of a star at the end of its life. The latter
objects are
now believed to result from outbursts of luminous, massive stars such
as the
class of luminous blue variables (LBVs).  Historical examples include
SN 1954J
in NGC 2403 (Humphreys \& Davidson 1994; Smith, Humphreys, \& Gehrz
2001), SN 1961V
in NGC
1058 (Goodrich et al.  1989; Filippenko et al. 1995; Van Dyk et al.
2002), and
SN 1997bs in NGC 3627 (Van Dyk et al. 2000).  An excellent review of
the
observational properties and challenges posed by LBVs is given by
Humphreys \&
Davidson (1994).

LBVs are evolved, massive, and unstable hot stars that undergo
sporadic,
sometimes violent, mass-loss events and whose nature is not well
understood.
The group is not entirely homogeneous since there are stars which are
luminous,
blue, and slightly variable that are not LBVs.  In general,
well-studied LBVs
have bolometric absolute magnitudes brighter than $-9.5$, suggesting
masses in
excess of 50 M$_\sun$. These stars exhibit relatively large mass-loss
rates of
10$^{-5}$ to 10$^{-4}$ M$_\sun$ yr$^{-1}$ which may be observed as an
infrared
excess from circumstellar ejecta.  An exceptional circumstellar
envelope is the
``homunculus'' surrounding $\eta$ Carinae.  LBVs exhibit eruptions of
1-2 mag
on time scales of 10--40 years and may have superoutbursts of more
than 2 mag
on time scales of hundreds to thousands of years. Their optical
spectra
typically show emission lines of H and \ion{He}{1}, as well as both
permitted
and forbidden lines of \ion{Fe}{2}.  Often these lines are accompanied
by
P-Cygni profiles with typical expansion velocities of 100--200 km
s$^{-1}$.  At
minimum light, the spectra are typical of hot supergiants with $T_{\sc
eff}
\approx 12,000--30,000$ K, while at maximum the spectra resemble
cooler
supergiants and are characteristic of an optically thick extended
photosphere
with $T_{\sc eff} \approx 7000-8000$ K.  Examples in the Galaxy
include P
Cygni, $\eta$ Carinae, S Doradus, and AG Carinae.  Their eruptions can
be
dramatic: $\eta$ Carinae underwent an outburst between 1830 and 1860
in which
it briefly reached $M_{\sc Bol} \approx -14$ mag.

In this paper we describe the discovery and evolution of a new
luminous variable star
designated
SN 2000ch (Papenkova \& Li 2000; Wagner et al. 2000a,b) which might be
related to LBVs.  The object
exhibits
erratic photometric variations on relatively fast time scales compared
to classical LBVs, but it has similar optical spectra when compared to
some LBVs in outburst.  During the
outburst,
its luminosity and size appear comparable to that of $\eta$ Carinae,
while in
quiescence it remains a luminous object and sufficiently bright for
detailed
study.

\section{Discovery and Identification}

Papenkova \& Li (2000), representing the LOSS team, discovered a new
variable
star in the field of the galaxy NGC 3432 on 2000 May 3.2 with the
0.76~m
Katzman Automatic Imaging Telescope (KAIT). The presence of the object
was
confirmed on unfiltered images obtained on May 4.2 (Figure 1).
Photometry of
the discovery images gives unfiltered magnitudes of 17.4 and 18.2 on
May 3 and
4, respectively. The position of the object is $\alpha$ =
10$^h$52$^m$41$\fs$40, $\delta$ = +36$\degr$40$\arcmin$08$\farcs$5
(J2000.0;
Papenkova \& Li 2000), which is 123$\arcsec$ east and 180$\arcsec$
north of the
diffuse nucleus of NGC 3432. An image obtained with KAIT on April 29.2
showed
nothing at the position of the new object brighter than a limiting
magnitude of
about 19. An identification chart for SN 2000ch is shown in Figure 2.

NGC 3432 is a diffuse and possibly interacting galaxy. Schwarzkopf \&
Dettmar
(2000a, b) recently studied NGC 3432 as part of a sample of
high-inclination galaxies that exhibit perturbations in their vertical
disk
structure as the result of tidal interactions and mergers.  Its
cataloged
heliocentric velocity from the NASA/IPAC Extragalactic Database is 616
km
s$^{-1}$, which corresponds to a distance of $10.5$ Mpc corrected for
Virgocentric flow and assuming $H_0 = 75$ km s$^{-1}$ Mpc$^{-1}$
(Schwarzkopf
\& Dettmar 2000b). Its revised Hubble type is 9.0, based on the
NASA/IPAC
Extragalactic Database, placing it between the Sc-Irr and Irr-I
galaxies of the
original Hubble (1926) classification scheme.

\section{Observations}

\subsection{Photometry}

Photometry of SN 2000ch was obtained with the 1.0~m
Ritchey-Chr\'{e}tien
reflector of the United States Naval Observatory Flagstaff Station
(USNOFS)
between 2000 May 7 and July 4.  The 1K $\times$ 1K CCD camera was used
in
combination with $UBVRI$ filters to image a $11\farcm3 \times
11\farcm3$ region
of the sky at a scale of $0\farcs67$ pixel$^{-1}$.  The seeing during
the run
typically averaged $\sim2\arcsec$.  Bias and twilight sky flatfield
frames were
also obtained to facilitate the data reduction, which was performed
using the
IRAF\footnote{IRAF is distributed by the National Optical Astronomy
Observatories, which are operated by the Association of Universities
for
Research in Astronomy, Inc., under cooperative agreement with the
National
Science Foundation.} package.

In addition, infrared (IR) images of NGC 3432 were obtained between
2000 May 13
and May 22 with the USNO 1.55~m Kaj Strand astrometric reflector and
the $256
\times 256$ pixel Rockwell HgCdTe (NICMOS) infrared camera (IRCAM).
$JHK$
images were obtained covering a field of view of $2\farcm3 \times
2\farcm3$ at
a scale of $0\farcs54$ pixel$^{-1}$.

$BVR$ images obtained of NGC 3432 and various standard stars under
photometric
conditions was used to accurately calibrate the magnitudes and colors
of eight
stars in the field of the galaxy and surrounding SN 2000ch.
Astrometry and
photometry for these stars are presented in Table 1 and their
identification
with respect to NGC 3432 and SN 2000ch is shown in Figure 2.

The brightness of SN 2000ch was measured by digital aperture
photometry of the
CCD images using the APPHOT package in IRAF and compared
differentially with
stars from the standard calibration of the surrounding field described
above.

Additional photometric observations of SN2000ch were obtained with
the 0.57-meter f/5.2 reflector of the Kle\v{t} Observatory
equipped with a SBIG ST-8 CCD camera (Tich\'a, Tich\'y, \& Moravec
2000). This CCD camera is equipped with Kodak KAF1600 detector
with quantum efficiency of about 40 per cent for a wavelength
interval 5000 - 8500 {\AA}. The field of view is $16.0 \times
10.7$ arcminutes. The limiting magnitude is $m_{V} = 19.5$ for a
120-second exposure under standard conditions. Under very good sky
conditions and no moonlight the limit is $m_{V} = 20.2$ for a
120-second exposure and $m_{V} = 20.4$ for a 180-second exposure.

The photometric measurements of SN 2000ch are summarized in Table 2.

\subsection{Spectroscopy}

Spectra of SN 2000ch were obtained at several observatories; a journal
of
observations is given in Table 3.  Spectral reductions were performed
in the
standard way using the IRAF package; these generally included bias
subtraction,
flat-fielding, wavelength calibration, flux calibration, and removal
of
telluric bands (e.g., Matheson et al. 2000). Mean extinction
coefficients were
assumed at each observatory. The spectra were placed on an absolute
flux scale
by matching the $V$-band magnitude of SN 2000ch inferred from the
interpolated
light curve shown in Figure 3.

The initial spectrum was obtained on 2000 May 6, the day following the
discovery announcement, using the CCD Spectropolarimeter (Schmidt,
Stockman, \&
Smith 1992) attached to the Steward Observatory 2.3~m Bok telescope at
Kitt
Peak National Observatory (KPNO). Standard-star observations were not
obtained
with the adopted grating, hence secondary calibration to absolute
units was
made with respect to white dwarfs observed for another program on both
that
night and the previous nights.  Removal of the atmospheric ``B" band
was not
possible.

Another spectrum was obtained on 2000 June 5.15 using the Ohio State
University
Boller and Chivens CCD spectrograph on the Hiltner 2.4-m telescope of
the MDM
Observatory on Kitt Peak.

SN 2000ch was observed twice with the Kast spectrograph (Miller \&
Stone 1993)
on the Shane 3-m reflector at Lick Observatory, on 2000 May 31.3 and
June
27.2. The airmass was 1.9--2.1 during the latter observation, and the
slit
position angle (68$^\circ$, through the star that is east-northeast of
the SN;
see Figure 2) was very close to the parallactic angle to avoid loss of
light at
the slit (Filippenko 1982).

On 2001 January 19, a final spectrum of SN 2000ch was obtained with
the Boller
and Chivens CCD spectrograph at the 2.3~m telescope at KPNO. The
absolute
intensity of the spectrum was scaled to match the average $V$
magnitude of
SN 2000ch in quiescence.

\section{Results}

\subsection{Archival Photometric Behavior}

Both POSS-I and POSS-II epoch frames were examined for additional
outbursts or
other activity related to SN 2000ch.  NGC 3432 appears on POSS-I 103aO
and
103aE plates obtained on 1953 May 8.  We find at the location of SN
2000ch
that $B > 20.1$ and $R > 20.0$ mag.
NGC 3432 was photographed again on four epochs as part of the POSS-II
survey.
We have examined these scanned images and find that for SN 2000ch on
1996 March
19, $B > 20$; 1998 April 17, $I = 19.8 \pm 0.5$; 1998 May 16, $R =
19.5 \pm
0.3$; and 1998 May 17, $R = 20.0 \pm 0.3$ mag.  These magnitudes were
determined by transforming the USNO-A2.0 instrumental
magnitudes to the standard system utilizing our photometric
calibration of the surrounding field as described above.  Our results
are in
good
agreement with an examination of the POSS images by Yamaoka (2000).
Thus, no
major or extended outbursts were detected in the rather limited sky
survey
plate material.

In their study of irregular galaxies, Schwarzkopf \& Dettmer (2000a)
imaged NGC
3432 in the $R$ band on 1997 May 31, using the 1.07~m Hall telescope
of the
Lowell Observatory on Anderson Mesa and the TI 800 $\times$ 800 pixel
CCD
camera. The image scale was $0\farcs36$ per pixel. The transparency
was
excellent and the seeing was $1\farcs1$.  SN 2002ch was easily
detected in
their image.  Photometry of this image gives $R = 19.2 \pm 0.1$ mag,
which is
comparable to the other quiescent magnitude determinations.

In addition, NGC 3432 was observed 32 times between 1998 December 11
and 2000
April 29 as part of the LOSS program.  SN 2000ch was typically fainter
than $R
= 19$ mag for nearly 90\% of these observations and fainter than $R =
18.4$ mag
for the remainder.  Taken together, examination of archival plate and
CCD image
material indicates that no previous outbursts comparable to that
discovered in
2000 May have been observed and that the data are consistent with a
roughly
constant pre-outburst brightness of $R = 19.4 \pm 0.4$ mag.  However,
previous
outbursts could certainly have been missed if their time scale was
similar to
that of this recent fast event.

\subsection{Light Curve}

An $R$-band light curve of SN 2000ch based on observations obtained
primarily
at the USNO and KAIT, with smaller contributions from other
observatories, is
shown in Figure 3. The unfiltered KAIT magnitudes have been
transformed to
approximate $R$ magnitudes based on the observed colors of the
variable; see
Riess (1999) for a discussion.  Archival observations obtained with
KAIT about
20 days before the May 3 outburst show SN 2000ch at $R = 19.3$ mag and
consistent with its quiescent brightness.  The object brightened
rapidly to $R
= 17.4$ mag on May 3.2, after which it faded quickly to $R = 20.8$ mag
on May
10. The object then recovered and brightened to $R = 18.6$ mag on May
14, after
which it varied between $R = 18.6$ and 19.4 mag to the conclusion of
our
observations on 2000 July 4.  The deep minimum may be indicative of a
brief
dust-formation phase, occultation, or eclipse, or more likely results
from the
evolution of an optically thick, cooling, ejected envelope.  If the
fluctuations observed after the deep minimum correspond to the
multiple
ejection and evolution of optically thick clouds or envelopes, then
the
evolutionary time scale for such events is $\sim$5--7 days.

The variability exhibited by the broad-band colors compared with the
V-band magnitude is shown in Figure 4 between TJD 51679-51706 and
immediately after the outburst.  The V-band magnitude monitors
changes in the overall brightness of the optical continuum while \bv\
is largely sensitive to changes in the slope of the continuum.  \ub\
is sensitive to changes in the strength of the Balmer continuum while
\vr\ and $R-I$ together are sensitive to changes in H$\alpha$
emission line flux.  Little variability is apparent in the \ub\ and
\bv\ light curves except for one outlying point near TJD 51684
suggesting little change in the slope of the optical continuum or the
strength of the Balmer continuum. Larger fluctuations are however
present in both \vr\ and $R-I$.  There is marginal evidence that the
variations are anticorrelated which suggests variability of H$\alpha$
emission on nightly timescales.

\subsection{Spectroscopy}

Spectra of SN 2000ch obtained on 2000 May 6, May 31, June 5, and
June 27 are shown in Figure 5. They exhibit prominent emission
lines of H$\alpha$, H$\beta$, and H$\gamma$ superposed on a flat
continuum. Weak He~I $\lambda$5876 and He~II $\lambda$4686
emission are also visible in spectrum (b) of Figure 5. The mean
full-width at half-maximum intensity (FWHM) of H$\alpha$ emission
is 1550 km s$^{-1}$ after correction for instrumental resolution.
This gives a mean expansion velocity of $\sim 800$ km s$^{-1}$ for
the envelope. The measured FWHM is comparable to other LBV and
$\eta$ Car variables.  A red wing is sometimes visible in the
H$\alpha$ profile. No forbidden lines are present in any of the
spectra, indicating that the electron density is in excess of
$\sim 10^8$ cm$^{-3}$; the absence of [O~I] emission suggests that
$n_e \ga 10^{9-10}$ cm$^{-3}$.  Our spectra lack the presence of
Fe II and [Fe II] emission lines that are typically present in the
spectra of LBVs in both outburst and quiescence.  In addition, our
spectra lack emission lines with accompanying P Cygni profiles
often seen in LBVs at visual maximum when the optically thick
expanded atmosphere resembles a much cooler A or F supergiant
(Humphereys and Davidson 1994).

A spectrum of SN 2000ch obtained 9 months after the 2000 outburst
on 2001 January 19, when the variable was fainter than 19 mag, is
shown in Figure 6. It is qualitatively similar to the previous
spectra and exhibits strong and broad Balmer emission lines with a
red wing, superposed on a slightly blue continuum. The FWHM of
H$\alpha$ emission is 1350 km s$^{-1}$, implying an expansion
velocity of $\sim 700$ km s$^{-1}$, comparable to that measured
during the outburst.  In general, we find that the optical
spectral properties of SN2000ch are similar during the outburst
and in quiescence.  In addition, no evidence is found in our
longslit spectroscopy obtained during the outburst or in
quiescence for narrow Balmer or nebular emission lines from an
underlying \ion{H}{2} region.

The H$\alpha$ to H$\beta$ intensity ratio increases significantly from
a value
of 3.4 on May 6, to 5.4 on May 31, and to 6.2 on June 27.  The $V-R$
color
increases from 0.21 to 0.88 mag, while $B-V$ varies between 0.24 and
0.29 mag
over approximately the same time interval, suggesting that all the
variation is
due to the increasing intensity of the emission lines, particularly
H$\alpha$.
This is readily apparent in Figure 7, where the H$\alpha$ line profile
observed
on 2000 May 31 is $\sim$50\% stronger than that obtained on 2001
January 19. A
possible interpretation is an increasing optical thickness in the
Balmer lines
(Case C recombination), as has been observed in some SNe~II such as SN
1987A
(Xu et al. 1992).

An estimate of the size of the line-emitting region can be derived
from the
luminosity in H$\alpha$ and recombination theory.  Assuming a distance
of 10.5
Mpc to NGC 3432 and an average integrated flux in H$\alpha$ of $2
\times
10^{-14}$ erg cm$^{-2}$ s$^{-1}$, the luminosity of H$\alpha$ is $2.5
\times
10^{38}$ erg s$^{-1}$. We can use the luminosity of the emission line
to
estimate the size of the emission region, since
$$ L({\rm H}\alpha) = (4/3) \pi R^3 N_H^2 \alpha_B h \nu , $$
where $R$ is the radius, $N_H$ is the hydrogen density, $\alpha_B$ is
the line
emissivity (Case C, $\sim10^{-24}$ erg cm$^3$ s$^{-1}$; Xu et al.
1992), and
$\nu$ is the frequency corresponding to H$\alpha$.  For an assumed
density of
10$^{10}$ cm$^{-3}$, we find that $R({\rm H}\alpha) \approx 0.2$ pc,
and
comparable in size to the famous ``homunculus'' surrounding $\eta$
Carinae.

\subsection{Spectral Energy Distribution and Maximum Luminosity}

The observed spectral energy distribution (SED) of SN 2000ch is shown
in Figure
8; it was constructed from optical photometry obtained on 2000 May 14
using the
USNO 1~m telescope and IR photometry obtained with the 1.55~m
telescope. The
spectrum consists of a hot component that gives rise to the blue
continuum and a marginally significant excess in the U-band, possibly
attributable to Balmer continuum emission, as
well as a cool component that is responsible for the apparent excess
in the K
band.  Accordingly, we have modeled the SED with two blackbodies in an
iterative approach assuming A$_V$ = 0.  We find that the hot component
is characterized by
a color
temperature of $T_1=7800 \pm 100$~K while the cool component is
characterized
by a color temperature of $T_2=1000 \pm 200$~K.  Note however that the
two-temperature blackbody fit is not very good with significant
deviations at
optical wavelengths; thus, the formally derived temperatures may be
somewhat
misleading with artifically small quoted uncertainties.  The projected
solid
angles
based on the fits imply that the ratio of their radii is $R_2/R_1=35
\pm 1$.
For a distance of 10.5 Mpc, $R_1=(6.7 \pm 1.5) \times 10^{13}$ cm so
that the
radius of the cool component $\sim2.5 \times 10^{15}$ cm.

Stronger evidence for the presence of Balmer continnum emission
can be found by examining the position of the variable in a
color-color diagram (Figure 9) constructed from our broad-band
photometry.  It can be seen that the variable occupies a small
region in the diagram, although its brightness varies
considerably.  The intrinsic colors of eight blue variables in M31
and M33 plus $\eta$ Car, MWC 349, and S Dor lie close to the
blackbody line and are bluer than SN 2000ch (Humphreys 1978).  It
is important to note that SN 2000ch also lies far from the locus
of points defining normal supergiants and above the line defining
radiating blackbodies.  This strongly suggests the presence of
Balmer continuum emission.  The tight concentration of points
implies that the strength and shape of the Balmer jump does not
vary much during the brightness fluctuations. These conclusions
are not influenced by the presence of interstellar reddening. This
interpretation is also consistent with the observations that most,
but not all, known luminous blue variables have more negative
$(U-B)_0$ colors for a given $(B-V)_0$ than main sequence or
supergiant stars (Humphreys 1978) indicating the presence of
Balmer continuum emission.

At its brightest, SN 2000ch reached an apparent unfiltered magnitude
of 17.4,
corresponding to an absolute magnitude of $-12.7$ at the adopted
distance of
NGC 3432, 10.5 Mpc ($\mu$ = 30.1 mag).  Its mean quiescent level of $R
= 19.4$
mag implies a persistent absolute magnitude of $-10.7$. Thus, even in
quiescence, SN 2000ch is an extremely luminous object.  It bears many
similarities to $\eta$ Carinae during its last major eruption between
1830 and
1860 when it reached $M_{\sc Bol} \approx -14$ mag, and was comparable
in size
to SN 2000ch. If the quiescent luminosity of SN 2000ch is constrained
by the
Eddington limit, the mass of the progenitor must exceed 40 M$_\sun$.

\section{Discussion}

The spectral features observed early in the outburst of SN 2000ch are
reminiscent of those observed in some M31 classical novae.  This
suggested to
Wagner et al. (2000a) that SN 2000ch might be related to a class of
super-bright
classical novae (Della Valle 1991), which have peak absolute
magnitudes $\sim$1
mag brighter than typical Galactic and M31 novae ($M_V \approx -9$
mag).
However, a bright-nova interpretation is inconsistent with the high
luminosity
($10^{6-7}\ L_\sun$), the relatively low expansion velocity, and the
fast
timescale that we observed. In the study presented by Della Valle
(1991), no
classical nova has been observed in the Galaxy, M31, or the Virgo
Cluster with
a luminosity comparable to that of SN 2000ch.

Based on the early data presented here, Filippenko (2000) suggested
that the
variable star in NGC 3432 could be a subluminous SN~IIn (Schlegel
1990;
Filippenko 1991) and thus perhaps related to the class of luminous
blue
variables (LBVs). Nearly all SNe IIn show hydrogen emission lines
consisting of
a broad complex base and a strong, narrow, and slightly blueshifted
component
superposed on a blue continuum.  Broad P-Cygni absorption components
are not
present (Schlegel 1990).  In addition, some SNe IIn have been observed
in
spiral galaxies in and near \ion{H}{2} regions or large massive
star-forming
complexes.  Their absolute magnitudes at discovery are typically
$-17.5$ to
$-18.0$ and thus comparable to most SNe II (Schlegel 1990).  The light
curves
of SNe~IIn, however, are not well characterized.

Recently, it has become apparent that the SN~IIn class is quite
heterogeneous
(Filippenko 1997). It consists of objects which are certainly {\it
bona-fide}
SNe~II in which core collapse has led to the destruction of the star.
But as
additional objects have been discovered whose characteristics resemble
SNe IIn,
it appears that some are in fact not true core-collapse SN but are the
result
of superoutbursts of luminous, massive stars.  Some of these stars are
luminous blue
variables on their way to becoming Wolf-Rayet stars. Objects in this
class
include $\eta$ Carinae (Davidson \& Humphreys 1997), SN 1954J in NGC
2403
(Humphreys \& Davidson 1994), SN 1961V in NGC 1058 (Goodrich et al.
1989;
Filippenko et al. 1995; Van Dyk et al. 2002), SN 1997bs in NGC 3627
(Van Dyk et
al.  2000), SN 1999bw in NGC 3198 (Filippenko, Li, \& Modjaz 1999),
and SN
2002kg in NGC 2403 (Schwartz, Filippenko, \& Chornock 2003).  All
these objects are
subluminous
with respect to more normal SNe~IIn, but otherwise exhibit similar
spectral
characteristics, apart from the very high velocities seen in some
SNe~IIn
(e.g., SN 1988Z; Filippenko 1997).

SN 2000ch in NGC 3432 is clearly similar to these other subluminous
objects and
thus probably related to the LBV class.  Its spectrum is similar to
that of SN
1997bs (Van Dyk et al. 2000), although SN 2000ch lacked the strong
\ion{Fe}{2}
emission observed in SN 1997bs.  The Balmer line widths are similar in
the two
objects, implying comparable outflow velocities.  In addition, the
maximum
absolute magnitude of $M_V \approx -13.8$ mag for SN 1997bs compares
reasonably
well with the value of $M_V = -12.7$ mag for SN 2000ch. While their
basic spectral
properties and maximum luminosities are similar, their host
environments are
quite different. SN 2000ch is located in an \ion{H}{2} region or large
star-forming complex at the edge of NGC 3432 (see Figure 2) while SN
1997bs
does not appear to reside in a comparable environment (Van Dyk et al.
2000).

The most striking difference is in their light curves.  SN 2000ch was
a very
fast, erratic event characterized by a rapid rise of $\sim2$ mag from
quiescence, a
local
minimum lasting $\sim$15~d, and finally a return back to quiescence.
In
contrast, the light curve of SN 1997bs reveals a gradual decline of
$\sim$6 mag
over $\sim$260~d. Smith, Humphreys, \& Gehrz (2001) point out that the
fast
photometric
variations exhibited by SN 2000ch are reminiscent of the oscillations
in
brightness reported for SN 1954J and $\eta$ Carinae prior to and
during their
giant eruptions.

In summary, SN 2000ch appears to be related to the LBV phenomenom but
with rapid photometric
variations
near maximum brightness.  While SN 1997bs has faded below
detectability in NGC
3627, SN 2000ch is quite luminous in quiescence ($R \approx 19.4$ mag)
and can
be studied in detail using modern instruments on large telescopes.  It
is
brighter than the progenitor of SN 1954J (Smith et al. 2001) and thus
can be
photometrically monitored frequently with small telescopes or observed
spectroscopically at relatively high spectral resolution.  We
encourage further monitoring and detailed
observations of this unusual outburst in an effort to discern its
nature and similarity to the class of LBVs.

\acknowledgments

The authors would like to thank Drs. Rodolfo Barba, Roberta Humphreys,
and Schuyler Van Dyk for their comments and suggestions regarding the
nature of SN 2000ch.  We are grateful for the assistance of the
support staff at KPNO,
Steward
Observotory, Lick Observatory, the MDM Observatory, and the United
States Naval
Observatory Flagstaff Station. R.M.W. acknowledges support from NASA
grants
to The
Ohio State University.  G.D.S. is grateful for NSF grant AST-9730792
to the
University of Arizona.  S.S. received partial support from NSF and
NASA grants
to Arizona State University.  The work of A.V.F.'s group at U. C.
Berkeley is
financed by National Science Foundation grant AST-9987438, as well as
by the
Sylvia and Jim Katzman Foundation.  KAIT was made possible by generous
donations from Sun Microsystems, Inc., the Hewlett-Packard Company,
AutoScope
Corporation, Lick Observatory, the National Science Foundation, the
University
of California, and the Katzman Foundation.




\clearpage

\begin{deluxetable}{rccrrrrrr}
\tabletypesize{\scriptsize} \tablecaption{Comparison Star
Magnitudes and Colors} \tablewidth{0pt} \tablehead{\colhead{Star}
& \colhead{RA (J2000)} & \colhead{Dec (J2000)} & \colhead{V} &
\colhead{$\sigma_{\sc V}$} & \colhead{B-V} & \colhead{$\sigma_{\sc
(B-V)}$} & \colhead{V-R} & \colhead{$\sigma_{\sc (V-R)}$} \\
\colhead{} & \colhead{hh mm ss.ss} & \colhead{$\pm$dd mm ss.s} &
\colhead{} & \colhead{} & \colhead{} & \colhead{} & \colhead{} &
\colhead{}}

\startdata

A & 10 52 52.36 & +36 38 23.2 & 14.534 & 0.002 & 1.453 &
   0.004 &     0.966 &   0.002 \\
B & 10 52 55.09 & +36 42 42.7 &      15.533 & 0.003 & 0.623 &
   0.006 &     0.342 &   0.004 \\
C & 10 52 47.49 & +36 41 42.7 &      16.002 &   0.004 & 0.844 &
   0.009 &     0.479 &   0.006 \\
D & 10 52 27.59 & +36 40 22.4 &      14.211 &   0.001 & 0.486 &
   0.002 &     0.287 &   0.001 \\
E & 10 52 50.23 & +36 37 38.8 &      16.659 &   0.007 & 0.645 &
    0.014 &     0.380 &   0.010 \\
F & 10 52 43.75 & +36 37 42.3 &      17.728 &   0.020 & 1.272 &
    0.061 &     0.784 &    0.023 \\
G & 10 52 49.25 & +36 42 52.2 &      17.004 &   0.009 & 0.602 &
    0.017 &     0.341 &    0.013 \\
H & 10 52 40.72 & +36 38  01.5 &      16.926 &   0.009 & 1.197 &
    0.028 &     0.760 &    0.011 \\

\enddata

\end{deluxetable}

\clearpage


\begin{deluxetable}{lcrrrrrrrrrrrrrrrr}
\tabletypesize{\scriptsize}
\setlength{\tabcolsep}{0.025in}
\tablecaption{Photometry of SN 2000ch}
\tablewidth{6.1in}
\tablehead{ \colhead{UT Date} &
\colhead{Observatory} & \colhead{U} & \colhead{$\sigma_{\sc U}$} &
\colhead{B} & \colhead{$\sigma_{\sc B}$} & \colhead{V} &
\colhead{$\sigma_{\sc V}$} & \colhead{R} & \colhead{$\sigma_{\sc
R}$} & \colhead{I} & \colhead{$\sigma_{\sc I}$} & \colhead{J} &
\colhead{$\sigma_{\sc J}$} & \colhead{H} & \colhead{$\sigma_{\sc
H}$} & \colhead{K} & \colhead{$\sigma_{\sc K}$} }

\startdata

10-APR-00   &   KAIT    &   \nodata &   \nodata &   \nodata &
\nodata &
\nodata &   \nodata &   19.2\tablenotemark{*}   &   0.4 &   \nodata &
\nodata &
 \nodata &   \nodata &   \nodata &   \nodata &   \nodata &   \nodata
\\
19-APR-00   &   KAIT    &   \nodata &   \nodata &   \nodata &
\nodata &
\nodata &   \nodata &   19.2\tablenotemark{*}   &   0.5 &   \nodata &
\nodata &
 \nodata &   \nodata &   \nodata &   \nodata &   \nodata &   \nodata
\\
24-APR-00   &   KAIT    &   \nodata &   \nodata &   \nodata &
\nodata &
\nodata &   \nodata &   19.5\tablenotemark{*}   &   0.3 &   \nodata &
\nodata &
 \nodata &   \nodata &   \nodata &   \nodata &   \nodata &   \nodata
\\
29-APR-00   &   KAIT    &   \nodata &   \nodata &   \nodata &
\nodata &
\nodata &   \nodata &   $>$19.2\tablenotemark{*}  &   \nodata &
\nodata &
\nodata &   \nodata &   \nodata &   \nodata &   \nodata &   \nodata &
\nodata \\
03-MAY-00   &   KAIT    &   \nodata &   \nodata &   \nodata &
\nodata &
\nodata &   \nodata &   17.4\tablenotemark{*}   &   0.1 &   \nodata &
\nodata &
 \nodata &   \nodata &   \nodata &   \nodata &   \nodata &   \nodata
\\
04-MAY-00   &   KAIT    &   \nodata &   \nodata &   \nodata &
\nodata &
\nodata &   \nodata &   18.2\tablenotemark{*}   &   0.1 &   \nodata &
\nodata &
 \nodata &   \nodata &   \nodata &   \nodata &   \nodata &   \nodata
\\
05-MAY-00   &   TLS &   \nodata &   \nodata &   \nodata &   \nodata &
\nodata &
 \nodata &   \nodata &   \nodata &   19.96   &   0.23    &   \nodata &
\nodata &
  \nodata &   \nodata &   \nodata &   \nodata \\
05-MAY-00   &   KLET    &   \nodata &   \nodata &   \nodata &
\nodata &
\nodata &   \nodata &   20.3\tablenotemark{*}   &   0.5 &   \nodata &
\nodata &
 \nodata &   \nodata &   \nodata &   \nodata &   \nodata &   \nodata
\\
06-MAY-00   &   KAIT    &   \nodata &   \nodata &   \nodata &
\nodata &
\nodata &   \nodata &   $>$18.9\tablenotemark{*}  &   \nodata &
\nodata &
\nodata &   \nodata &   \nodata &   \nodata &   \nodata &   \nodata &
\nodata \\
06-MAY-00   &   KLET    &   \nodata &   \nodata &   \nodata &
\nodata &
\nodata &   \nodata &   $>$20.5   &   \nodata &   \nodata &   \nodata
&   \nodata
&   \nodata &   \nodata &   \nodata &   \nodata &   \nodata \\
07-MAY-00   &   USNO    &   \nodata &   \nodata &   \nodata &
\nodata &
\nodata &   \nodata &   20.31   &   0.08    &   \nodata &   \nodata &
\nodata &
 \nodata &   \nodata &   \nodata &   \nodata &   \nodata \\
09-MAY-00   &   USNO    &   \nodata &   \nodata &   \nodata &
\nodata &
\nodata &   \nodata &   20.73   &   0.11    &   \nodata &   \nodata &
\nodata &
 \nodata &   \nodata &   \nodata &   \nodata &   \nodata \\
10-MAY-00   &   USNO    &   \nodata &   \nodata &   \nodata &
\nodata &
\nodata &   \nodata &   20.83   &   0.1 &   \nodata &   \nodata &
\nodata &
\nodata &   \nodata &   \nodata &   \nodata &   \nodata \\
11-MAY-00   &   USNO    &   \nodata &   \nodata &   \nodata &
\nodata &
\nodata &   \nodata &   20.48   &   0.08    &   \nodata &   \nodata &
\nodata &
 \nodata &   \nodata &   \nodata &   \nodata &   \nodata \\
12-MAY-00   &   USNO    &   \nodata &   \nodata &   \nodata &
\nodata &
\nodata &   \nodata &   19.64   &   0.07    &   \nodata &   \nodata &
\nodata &
 \nodata &   \nodata &   \nodata &   \nodata &   \nodata \\
13-MAY-00   &   USNO    &   \nodata &   \nodata &   \nodata &
\nodata &
\nodata &   \nodata &   19.76   &   0.06    &   19.19   &   0.04    &
19.25   &
 0.12    &   $>$18.0   &   \nodata &   $>$17.1   &   \nodata \\
14-MAY-00   &   USNO    &   18.42   &   0.07    &   19.12   &   0.05
&   18.83
 &   0.03    &   18.62   &   0.02    &   18.24   &   0.02    &   18.02
&   0.08
  &   17.65   &   0.15    &   16.7    &   0.16    \\
15-MAY-00   &   USNO    &   \nodata &   \nodata &   19.65   &   0.05
&   19.37
 &   0.06    &   18.84   &   0.03    &   18.77   &   0.07    &
\nodata &
\nodata &   \nodata &   \nodata &   \nodata &   \nodata \\
17-MAY-00   &   USNO    &   \nodata &   \nodata &   \nodata &
\nodata &
\nodata &   \nodata &   19.25   &   0.09    &   \nodata &   \nodata &
\nodata &
 \nodata &   \nodata &   \nodata &   \nodata &   \nodata \\
18-MAY-00   &   USNO    &   \nodata &   \nodata &   \nodata &
\nodata &   19.99
 &   0.1 &   19.36   &   0.04    &   \nodata &   \nodata &   18.61   &
0.14    &
  \nodata &   \nodata &   \nodata &   \nodata \\
19-MAY-00   &   USNO    &   \nodata &   \nodata &   19.52   &   0.05
&   19.56
 &   0.05    &   19.06   &   0.04    &   \nodata &   \nodata &   18.81
&   0.12
  &   18.11   &   0.11    &   $>$17.9   &   \nodata \\
20-MAY-00   &   USNO    &   \nodata &   \nodata &   \nodata &
\nodata &
\nodata &   \nodata &   19.08   &   0.07    &   \nodata &   \nodata &
\nodata &
 \nodata &   \nodata &   \nodata &   \nodata &   \nodata \\
21-MAY-00   &   USNO    &   19.36   &   0.11    &   20.01   &   0.05
&   19.79
 &   0.03    &   19.08   &   0.02    &   19.22   &   0.17    &   18.54
&   0.06
  &   18.2    &   0.2 &   $>$17.1   &   \nodata \\
22-MAY-00   &   USNO    &   19.14   &   0.05    &   19.65   &   0.07
&   19.49
 &   0.04    &   18.98   &   0.03    &   18.87   &   0.09    &   18.54
&   0.06
  &   \nodata &   \nodata &   \nodata &   \nodata \\
23-MAY-00   &   USNO    &   18.99   &   0.04    &   19.75   &   0.05
&   19.41
 &   0.04    &   18.95   &   0.03    &   18.84   &   0.04    &
\nodata &
\nodata &   $>$18.4    &   \nodata   &   \nodata &   \nodata \\
24-MAY-00   &   USNO    &   19.13   &   0.06    &   19.92   &   0.05
&   19.68
 &   0.04    &   19.06   &   0.03    &   \nodata &   \nodata &
\nodata &
\nodata &   \nodata &   \nodata &   \nodata &   \nodata \\
25-MAY-00   &   USNO    &   18.7    &   0.04    &   19.48   &   0.04
&   19.28
 &   0.04    &   18.81   &   0.03    &   18.76   &   0.06    &
\nodata &
\nodata &   \nodata &   \nodata &   \nodata &   \nodata \\
26-MAY-00   &   USNO    &   18.55   &   0.04    &   19.24   &   0.04
&   19.01
 &   0.02    &   18.58   &   0.02    &   18.48   &   0.04    &
\nodata &
\nodata &   \nodata &   \nodata &   \nodata &   \nodata \\
27-MAY-00   &   USNO    &   18.57   &   0.1 &   19.35   &   0.04    &
19.08   &
 0.04    &   18.61   &   0.02    &   18.58   &   0.06    &   \nodata &
\nodata &
  \nodata &   \nodata &   \nodata &   \nodata \\
28-MAY-00   &   USNO    &   18.94   &   0.04    &   19.62   &   0.03
&   19.38
 &   0.03    &   18.83   &   0.02    &   18.92   &   0.06    &
\nodata &
\nodata &   \nodata &   \nodata &   \nodata &   \nodata \\
29-MAY-00   &   USNO    &   19.2    &   0.05    &   19.95   &   0.04
&   19.68
 &   0.04    &   19.04   &   0.04    &   19.17   &   0.07    &
\nodata &
\nodata &   \nodata &   \nodata &   \nodata &   \nodata \\
30-MAY-00   &   USNO    &   19.34   &   0.06    &   20.08   &   0.06
&   19.85
 &   0.05    &   19.24   &   0.04    &   19.19   &   0.12    &
\nodata &
\nodata &   \nodata &   \nodata &   \nodata &   \nodata \\
31-MAY-00   &   USNO    &       &   \nodata &   20.02   &   0.05    &
19.78   &
 0.04    &   19.23   &   0.04    &   \nodata &   \nodata &   \nodata &
\nodata &
  \nodata &   \nodata &   \nodata &   \nodata \\
01-JUN-00   &   USNO    &   19.16   &   0.06    &   19.83   &   0.05
&   19.66
 &   0.05    &   19.1    &   0.03    &   19.04   &   0.05    &
\nodata &
\nodata &   \nodata &   \nodata &   \nodata &   \nodata \\
02-JUN-00   &   USNO    &   18.99   &   0.05    &   19.78   &   0.04
&   19.58
 &   0.05    &   19.04   &   0.03    &   18.97   &   0.05    &
\nodata &
\nodata &   \nodata &   \nodata &   \nodata &   \nodata \\
02-JUN-00   &   CA  &   \nodata &   \nodata &   19.87   &   0.05    &
19.56   &
 0.03    &   19.01   &   0.02    &   18.97   &   0.06    &   \nodata &
\nodata &
  \nodata &   \nodata &   \nodata &   \nodata \\
03-JUN-00   &   CA  &   \nodata &   \nodata &   \nodata &   \nodata &
\nodata &
 \nodata &   18.86   &   0.08    &   19.03   &   0.08    &   \nodata &
\nodata &
  \nodata &   \nodata &   \nodata &   \nodata \\
04-JUN-00   &   CA  &   \nodata &   \nodata &   19.74   &   0.07    &
19.46   &
 0.06    &   19.01   &   0.04    &   18.87   &   0.08    &   \nodata &
\nodata &
  \nodata &   \nodata &   \nodata &   \nodata \\
05-JUN-00   &   USNO    &   18.87   &   0.06    &   19.72   &   0.04
&   19.48
 &   0.04    &   18.91   &   0.03    &   18.91   &   0.09    &
\nodata &
\nodata &   \nodata &   \nodata &   \nodata &   \nodata \\
06-JUN-00   &   USNO    &   \nodata &   \nodata &   19.55   &   0.04
&   19.34
 &   0.04    &   18.91   &   0.03    &   \nodata &   \nodata &
\nodata &
\nodata &   \nodata &   \nodata &   \nodata &   \nodata \\
07-JUN-00   &   USNO    &   18.79   &   0.06    &   19.51   &   0.04
&   19.32
 &   0.04    &   18.81   &   0.03    &   18.86   &   0.06    &
\nodata &
\nodata &   \nodata &   \nodata &   \nodata &   \nodata \\
08-JUN-00   &   USNO    &   18.56   &   0.08    &   19.31   &   0.04
&   19.16
 &   0.04    &   18.67   &   0.02    &   18.88   &   0.1 &   \nodata &
\nodata &
  \nodata &   \nodata &   \nodata &   \nodata \\
09-JUN-00   &   USNO    &   \nodata &   \nodata &   19.25   &   0.05
&   19.11
 &   0.04    &   18.63   &   0.03    &   \nodata &   \nodata &
\nodata &
\nodata &   \nodata &   \nodata &   \nodata &   \nodata \\
10-JUN-00   &   USNO    &   \nodata &   \nodata &   19.28   &   0.04
&   19.02
 &   0.04    &   18.62   &   0.04    &   \nodata &   \nodata &
\nodata &
\nodata &   \nodata &   \nodata &   \nodata &   \nodata \\
27-JUN-00   &   USNO    &   \nodata &   \nodata &   \nodata &
\nodata &   20.23
 &   0.06    &   19.35   &   0.05    &   \nodata &   \nodata &
\nodata &
\nodata &   \nodata &   \nodata &   \nodata &   \nodata \\
02-JUL-00   &   USNO    &   \nodata &   \nodata &   \nodata &
\nodata &
\nodata &   \nodata &   19.37   &   0.03    &   \nodata &   \nodata &
\nodata &
 \nodata &   \nodata &   \nodata &   \nodata &   \nodata \\
04-JUL-00   &   USNO    &   \nodata &   \nodata &   \nodata &
\nodata &
\nodata &   \nodata &   19.33   &   0.05    &   \nodata &   \nodata &
\nodata &
 \nodata &   \nodata &   \nodata &   \nodata &   \nodata \\

\enddata
\tablenotetext{*}{Unfiltered KAIT Magnitudes}
\end{deluxetable}

\clearpage

\begin{deluxetable}{llcccccll}
\tablecaption{Journal of Spectroscopic Observations}
\tablehead{
\colhead{UT Date} & \colhead{Tel.\tablenotemark{a}}
&\colhead{Range\tablenotemark{b}}
&\colhead{Res.\tablenotemark{c}}& \colhead{Air.\tablenotemark{d}} &
\colhead{Slit}
& \colhead{Seeing} &\colhead{Exp.}
&\colhead{Observers\tablenotemark{e}} \\
& & (\AA) & (\AA) & & ($''$) & ($''$) & {~~(s)~~}&
}
\startdata
 2000-05-06 & S2.3p & 4320--7400 & 8 & 1.06 & 2.0 & 1.3 & 1200 & GDS,
PS \\
 2000-05-31 & L3.0 & 4250--6950 & 9 & 1.6 & 3.0 & 2 & 1500 & AF,AC \\
 2000-06-05 & M2.4 & 3900--7500 & 7.5 & 1.3 & 5.0 & 1.2 & 5400 & RMW
\\
 2000-06-27 & L3.0 & 4250--6950 & 9 & 2.0\tablenotemark{f} & 2.0 & 2 &
2700 & AF,AC \\
 2001-01-19 & S2.3s & 4300--7500 & 7.5 & 1.05 & 1.5 & 1.5 & 5400 & RMW
\\
\enddata
\tablenotetext{a}{S2.3p = Steward Observatory 2.3-m Bok telescope +
 CCD Spectropolarimeter; S2.3s = Steward Observatory 2.3-m Bok
telescope +
 Boller and Chivens CCD Spectrograph; M2.4 = MDM Observatory 2.4-m
Hiltner
 telescope + Boller and Chivens CCD Spectrograph; L3.0 = Lick 3-m
3.0-m
 Shane telescope + Kast CCD Spectrograph.}
\tablenotetext{b}{Observed wavelength range of spectrum. }
\tablenotetext{c}{Approximate resolution (FWHM) of spectrum. }
\tablenotetext{d}{Average airmass of observation.}
\tablenotetext{e}{GDS = G. Schmidt, PS = P. Smith, AF = A. Filippenko,
AC = A. Coil, RMW = R. M. Wagner}
\tablenotetext{f}{The slit was about $20^\circ$ from the parallactic
angle.
In all other cases, either the slit position angle was close to the
parallactic
angle, or the airmass is low.}

\end{deluxetable}


\clearpage



\figcaption{Unfiltered discovery images of the luminous variable star
(SN
2000ch)
in the field of NGC 3432 obtained with KAIT on 2000 May 3.2 {\it
(left)} and
May 4.2 {\it (right)}.  The
variable is indicated by the tickmarks.  Each frame measures 6\farcm1
$\times$ 6\farcm6 on a side.  North is to the top and east is to the
left.  Photometry of the variable gives 17.4 and 18.2
mag
respectively.}

\figcaption{Identification chart for the variable star in NGC 3432
obtained with
the USNO 1.0-m telescope.  Lettered field stars correspond to those
listed in
Table 1. North is to the top and east is to the left.  The image
covers 6$\arcmin$ $\times$ 6$\arcmin$ on a side.}

\figcaption{The $R$-band light curve of the variable star in NGC 3432
is
shown
beginning on 2000 Apr. 10, just prior to the discovery by KAIT on 2000
May 3,
and continuing through observations obtained on July 4.  The light
curve
consists of a rapid brightening to $R = 17.4$ mag followed by an
equally rapid
fading to $R = 20.8$ mag on May 10.  Subsequently, the object
brightened to $R
= 18.6$ mag on May 14 after which it varied between $R = 18.6$ and
19.4
mag. The KAIT and Klet ``open" observations are unfiltered, but
correspond
roughly to the $R$ band.}

\figcaption{Variation of the broad-band colors of SN 2000ch compared
with
the overall V-band light curve immediately after the outburst.  The
largest variations occur in \vr\ and $R-I$ with some evidence they
are anticorrelated suggesting variability of H$\alpha$ emission.}

\figcaption{Spectra of the luminous variable star (SN 2000ch) in NGC
3432
obtained
with the Steward 2.3-m (a), Lick 3-m (b and d), and MDM 2.4-m (c)
telescopes on
2000 May 6, May 31, Jun. 5, and Jun. 27 respectively. The observed
FWHM of
H$\alpha$ emission is 1350 km s$^{-1}$, giving an average expansion
velocity of
$\sim 700$ km s$^{-1}$.  The spectra show little evolution over the
interval of
our observations.}

\figcaption{Spectrum of the luminous variable star obtained with the
Steward
Observatory Bok 2.3-m telescope on 2001 January 19, 9 months after the
2000
outburst. This spectrum is similar to those obtained during the
outburst.}

\figcaption{H$\alpha$ line profiles obtained on 2000 May 31 {\it
(solid
line)} at
Lick Observatory and on 2001 January 19 {\it (dotted line)} at Steward
Observatory.  Note the large change in the strength of the line.}

\figcaption{The observed spectral energy distribution of SN 2000ch is
shown based
on USNO photometry obtained on 2000 May 14. The {\it dotted line} is a
7800~K
blackbody, while the {\it dashed line} is a 1000~K blackbody.  The
composite
model spectrum is shown by the {\it heavy solid line}.  There is
evidence for an IR
excess, possibly due to heated dust or free-free emission. The
marginally
significant excess in the U-band might be attributable to Balmer
continuum
emission.  The two-temperature blackbody fit is not particularly good,
but is
meant to illustrate only the general trends.}

\figcaption{Position of SN 2000ch during its outburst in a \ub\
versus \bv\
color-color plot.  The locus of points corresponding to main sequence
stars (MS: dotted line) and supergiants (SG: dashed line) is shown.
The blackbody line (BB) and a reddening vector corresponding to
$E(B-V)$ = 0.2 are also shown.  The variable lies above the line
defining radiative
blackbodies suggesting the presence of Balmer continuum emission.}



\begin{thebibliography}{}

\bibitem[Davidson \& Humphreys(1997)]{1997ARA&A..35....1D} Davidson,
K.,~\&
  Humphreys, R.~M.\ 1997, \araa, 35, 1
\bibitem[Della Valle(1991)]{1991A&A...252L...9D} Della Valle, M.\
1991,
  \aap, 252, L9
\bibitem[Filippenko(1982)]{1982PASP...94..715F} Filippenko, A.~V.\
1982,
  \pasp, 94, 715
\bibitem[Filippenko 1991]{avf91}
  Filippenko, A. V. 1991, in SN 1987A and Other Supernovae, ed. I. J.
Danziger \&
  K. Kj\"ar (Garching: ESO), 342
\bibitem[Filippenko(1997)]{1997ARA&A..35..309F} Filippenko, A.~V.\
1997,
  \araa, 35, 309
\bibitem[Filippenko(2000)]{2000IAUC.7421....3F} Filippenko, A.~V.\
2000,
  \iaucirc\ 7421
\bibitem[Filippenko, Li, \& Modjaz(1999)]{1999IAUC.7152....2F}
Filippenko,
  A.~V., Li, W.~D., \& Modjaz, M.\ 1999, \iaucirc\ 7152
\bibitem[Filippenko et al. 2001]{avf01} Filippenko, A. V., Li, W. D.,
  Treffers, R. R., \& Modjaz, M. 2001, in Small-Telescope Astronomy on
  Global Scales, ed. W. P. Chen, C. Lemme, and B. Paczynski (San
Francisco:
  ASP), 121
\bibitem[Filippenko et al. 1995]{avf95}
  Filippenko, A. V., et al. 1995, \aj, 110, 2261 (erratum 112,
806 [1996])
\bibitem[Goodrich et al. (1989)]{goo89} Goodrich, R. W., Stringfellow,
G. S.,
   Penrod, G. D., \& Filippenko, A. V. 1989, \apj, 342, 908
\bibitem[Hubble(1926)]{1926ApJ....64..321H} Hubble, E.~P.\ 1926, \apj,
64,
  321
\bibitem[Humphreys 1978]{rh78} Humphreys, R. M. 1978, \apj, 219, 445
\bibitem[Humphreys \& Davidson 1994]{rh94}
  Humphreys, R. M., \& Davidson, K. 1994, \pasp, 106, 1025
\bibitem[Li et al. 2000]{li00} Li, W. D., et al. 2000, in Cosmic
Explosions,
 ed. S. S. Holt \& W. W. Zhang (New York: AIP), 103
\bibitem[Matheson et al. 2000]{math00} Matheson, T., Filippenko, A.
V.,
  Ho, L. C., Barth, A. J., \& Leonard, D. C. 2000, \aj, 120, 1499
\bibitem[Miller \& Stone 1993]{jm93}
  Miller, J. S., \& Stone, R. P. S. 1993, Lick Obs. Tech. Rep. 66
(Santa
  Cruz: Lick Obs.)
\bibitem[Papenkova \& Li(2000)]{2000IAUC.7415....1P} Papenkova, M., \&
Li,
  W.~D.\ 2000, \iaucirc\ 7415
\bibitem[Riess et al. 1999]{rie99} Riess, A. G., et al. 1999, \aj,
118, 2675
\bibitem[Schlegel 1990]{ems90}
  Schlegel, E. M. 1990, \mnras, 244, 269
\bibitem[Schmidt, Stockman, \& Smith(1992)]{1992ApJ...398L..57S}
Schmidt,
  G.~D., Stockman, H.~S., \& Smith, P.~S.\ 1992, \apjl, 398, L57
\bibitem[Schwartz et al. 2003]{sch03} Schwartz, M., Li, W.,
Filippenko, A. V.,
  \& Chornock, R. 2003, \iaucirc\ 8051
\bibitem[Schwarzkopf \& Dettmar 2000a]{sd00a}
  Schwarzkopf, U., \& Dettmar, R. J. 2000a, \aaps, 144, 85
\bibitem[Schwarzkopf \& Dettmar(2000b)]{2000A&A...361..451S}
Schwarzkopf,
  U., \& Dettmar, R.-J.\ 2000b, \aap, 361, 451
\bibitem[Smith et al. 2001]{ns01}
  Smith, N., Humphreys, R. M., \& Gehrz, R. D. 2001, \pasp, 113, 692
\bibitem[Ticha et al. 2000]{ttm00}
  Tich\'a, J., Tich\'y, M., \& Moravec, Z. 2000, Planet.Space.Sci, 48,
787
\bibitem[Van Dyk et al. 2000]{svd00}
  Van Dyk, S. D., et al. 2000, \pasp, 112, 1532
\bibitem[Van Dyk et al. 2002]{svd02}
  Van Dyk, S. D., Filippenko, A. V., \& Li, W. 2002, \pasp, 114, 700
\bibitem[Wagner et al.(2000)]{2000IAUC.7417....2W} Wagner, R.~M.,
Schmidt,
  G.~D., Smith, P., Hines, D., \& Starrfield, S.~G.\ 2000a, \iaucirc\
7417
\bibitem[Wagner et al.(2000b)]{2000AAS...197.4413W} Wagner, R.~M., et
al.\
  2000b, \baas, 32, 1474
\bibitem[Xu, McCray, Oliva, \& Randich(1992)]{1992ApJ...386..181X} Xu,
Y.,
  McCray, R., Oliva, E., \& Randich, S.\ 1992, \apj, 386, 181
\bibitem[Yamaoka(2000)]{2000IAUC.7419....1Y} Yamaoka, H.\ 2000,
\iaucirc\
  7419

\end{thebibliography}
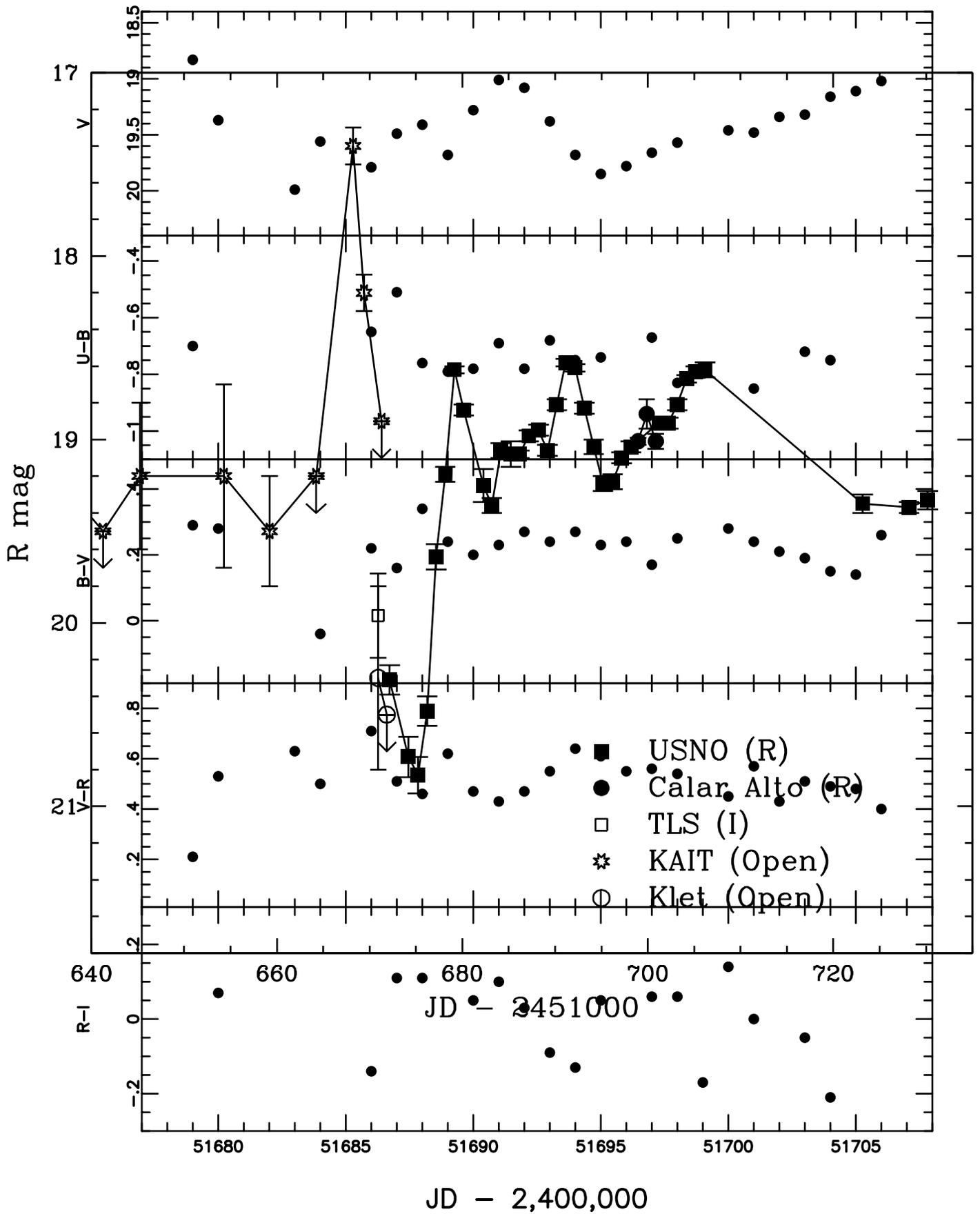
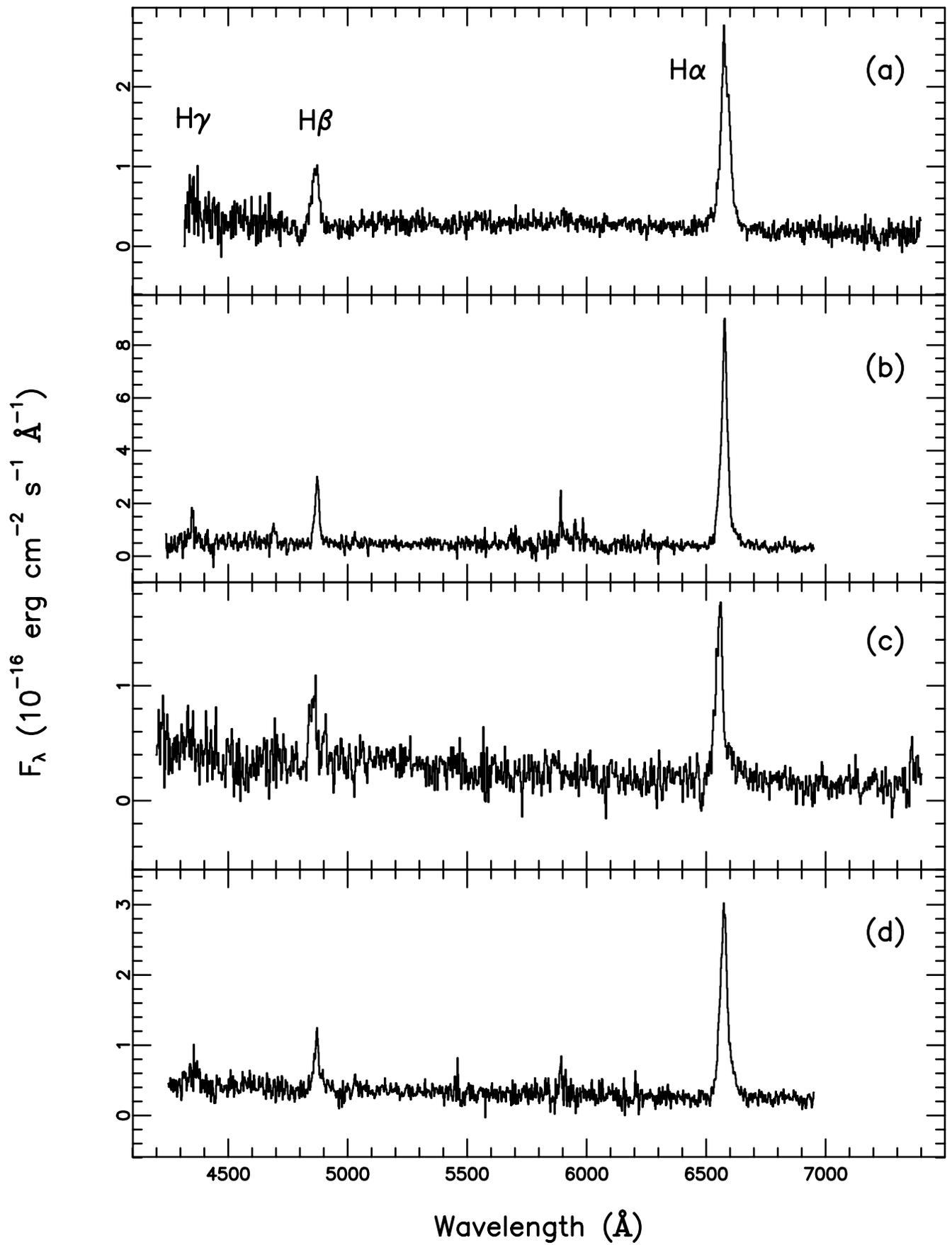
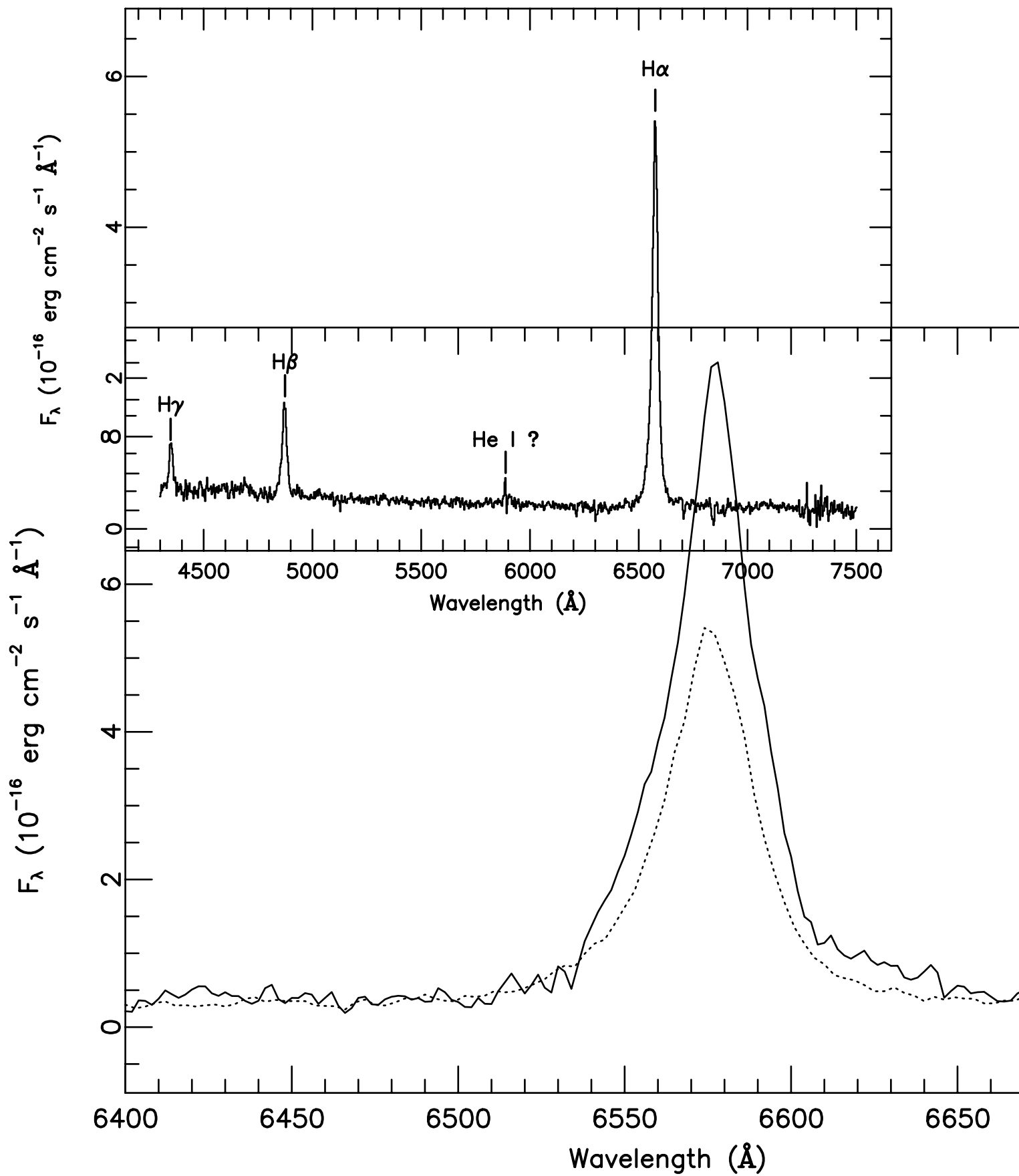
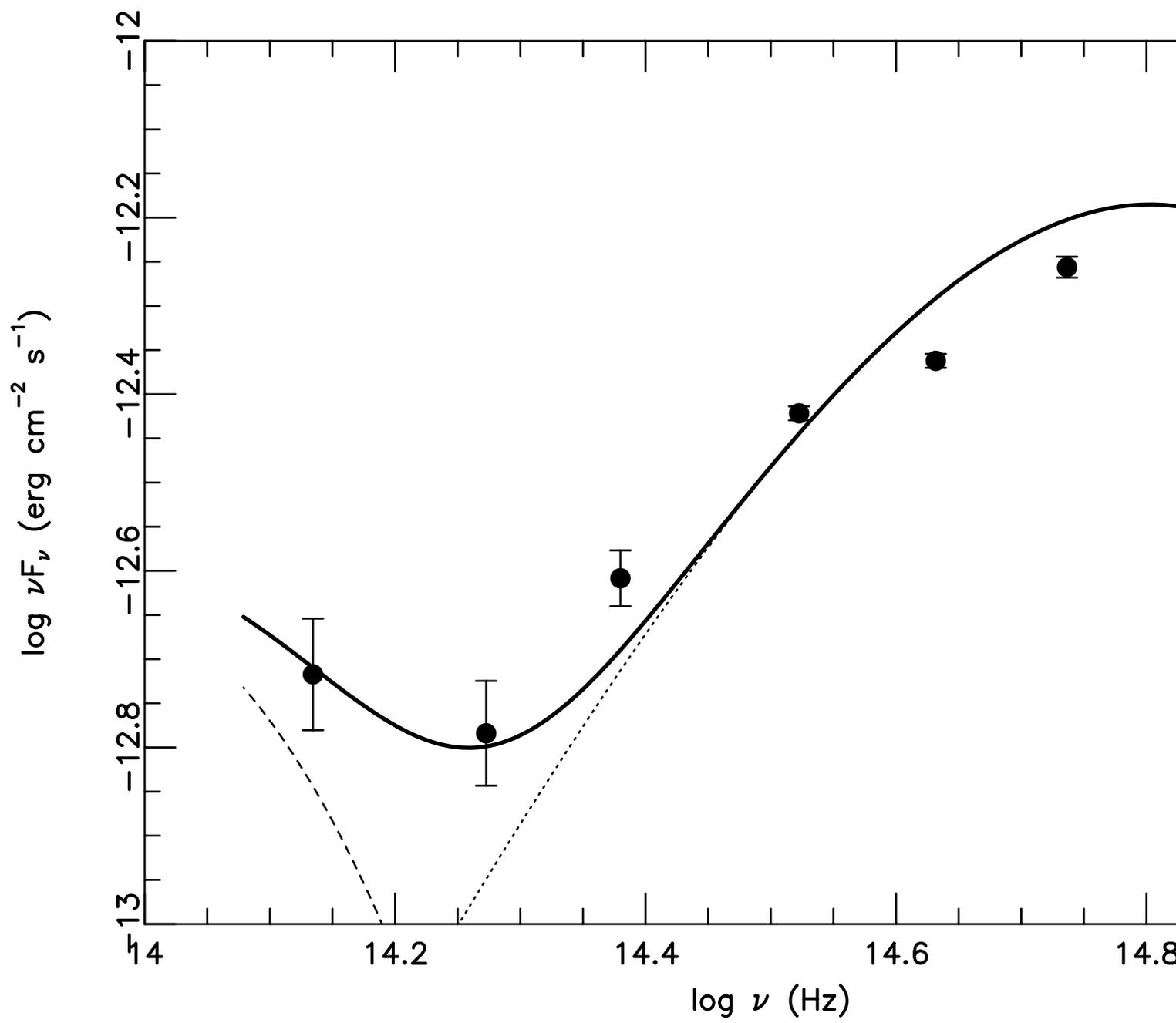
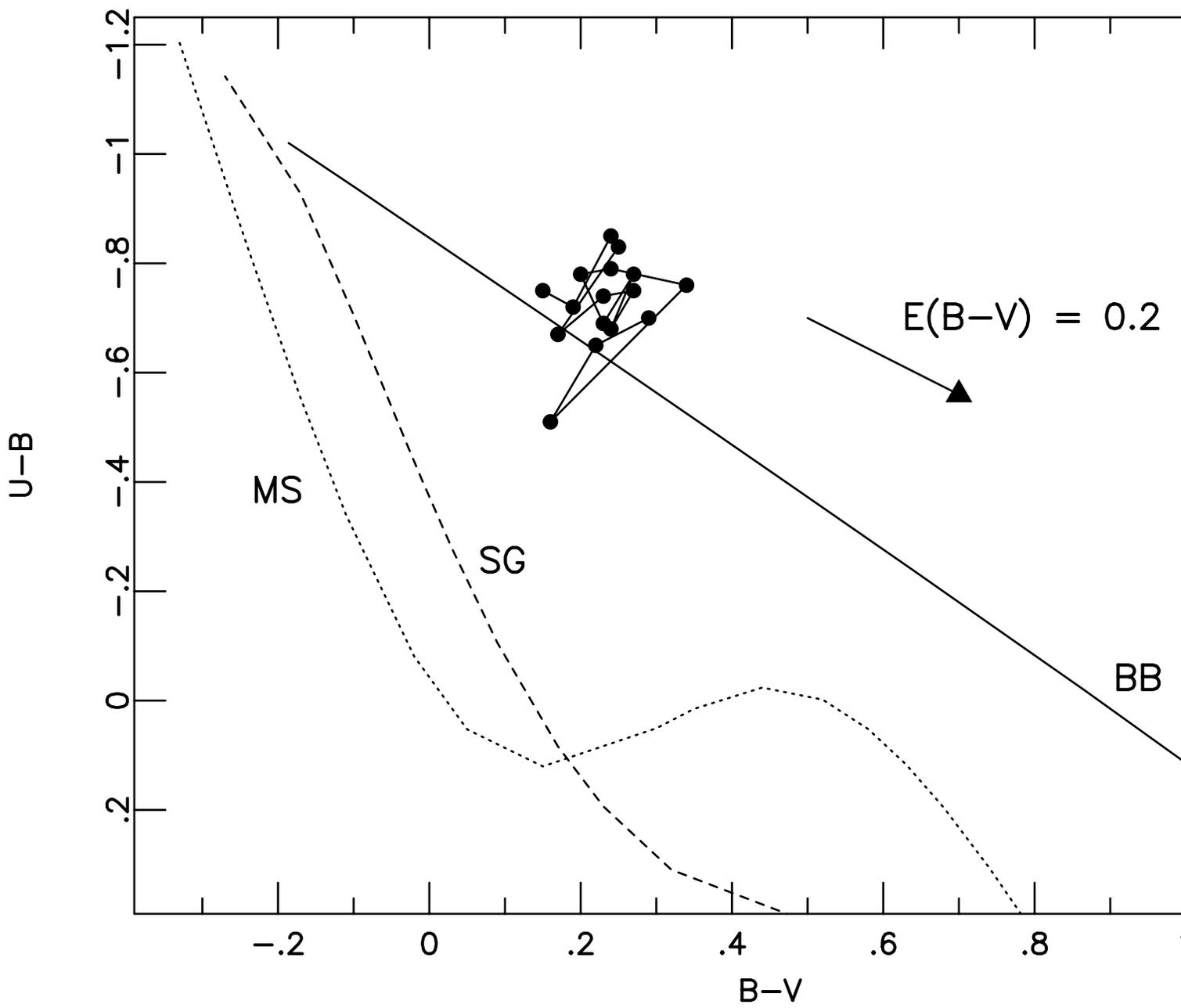
\end{document}